# Electric polarization of magnetic textures: new horizons of micromagnetism

A.P. Pyatakov[a,b]*, G.A. Meshkov[a], A.K. Zvezdin[b]

[a]M.V. Lomonosov Moscow State University, Leninskie gori, Moscow, 119991, Russia
[b]A.M. Prokhorov General Physics Institute, Vavilova St., 38 Moscow, 119991, Russia

**Abstract**

A common scenario of magnetoelectric coupling in multiferroics is the electric polarization induced by spatially modulated spin structures. It is shown in this paper that the same mechanism works in magnetic dielectrics with inhomogeneous magnetization distribution: the domain walls and magnetic vortexes can be the sources of electric polarization. The electric field driven magnetic domain wall motion is observed in iron garnet films. The electric field induced nucleation of vortex state of magnetic nanodots is theoretically predicted and numerically simulated. From the practical point of view the electric field control of micromagnetic structures is promising for applications in low-power-consumption spintronic and magnonic devices.

### 1. Introduction

It is a well known fact of classical electromagnetism that *temporally varying* magnetic and electric fields are coupled while static fields can be considered independently. However it is not generally known that the later statement is true while the magnetically ordered media are not considered. In the case of magnetics a *spatially varying* magnetic order parameter can induce electric polarization in the material due to the general phenomenon of flexoelectricity [1-6].

This is common scenario of magnetoelectric coupling in the so-called spiral multiferroics, i.e. the media whose ferroelectricity is induced by intrinsic spatially modulated spin structure [7-24]. In the same way the conventional magnetic textures like domain walls or vortexes can be the sources of electric polarization. This inhomogeneous magnetoelectric interaction (or spin flexoelectricity) is responsible for new phenomena in micromagnetism that are reviewed in this paper.

* Corresponding author. Tel.: +7-495-939-4138; fax: +7-495-932-8820.
E-mail address: pyatakov@physics.msu.ru

## 2. Spin flexoelectricity

What does prefix "flexo" mean in the "flexoelectricity" term? Let us consider the highly symmetrical case of a cubic crystal with a center of symmetry. In accordance with Curie principle mechanical deformations reduce the symmetry of the crystal (fig. 1). However only the bending, or *flexural* strain, with the bottom layers compressed and the top ones stretched, violates inversion symmetry and points out the vertical polar axis (fig. 1 c) that is the prerequisite to the electric polarization. The electric polarization induced by the flexural strain is regarded as *flexoelectricity*.

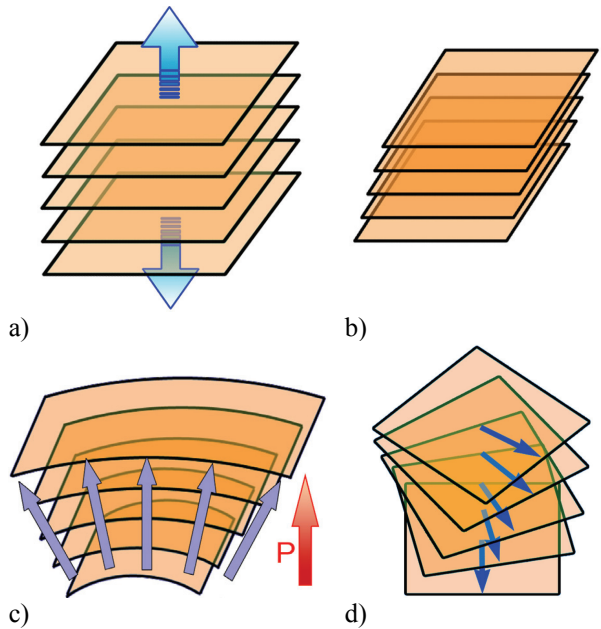

Figure 1. Four types of mechanical deformations a) longitudinal strain  b) shear strain c) flexural strain with vertically oriented strain gradient vector ) d) the twist strain.

In the case of magnetic media there is no need to bend crystal. Indeed, from the same symmetry arguments it follows that a spatially modulated spin structure of a cycloid type (fig. 2 a) induces ferroelectricity in analogy to the flexural deformation (while a helicoidal spiral similar to twisting deformation (fig. 2b) remains centrosymmetrical[2]).

Mathematically spin flexoelectricity is described by the term in the free energy proportional to the spatial derivatives of a magnetic order parameter [7]:

$$F_{ME} = -\gamma_{ijkl} \cdot P_i \cdot m_j \cdot \nabla_k m_l ,  \quad (1)$$

where $\mathbf{m}=\mathbf{M}(\mathbf{r})/M_s$ is the unit magnetization vector, $M_s$ is saturation magnetization, $\mathbf{P}$ is the electric polarization, $\nabla$ is the vector differential operator, $\gamma_{ijkl}$ is the tensor of the inhomogeneous magnetoelectric interaction that is determined by the symmetry of the crystal.

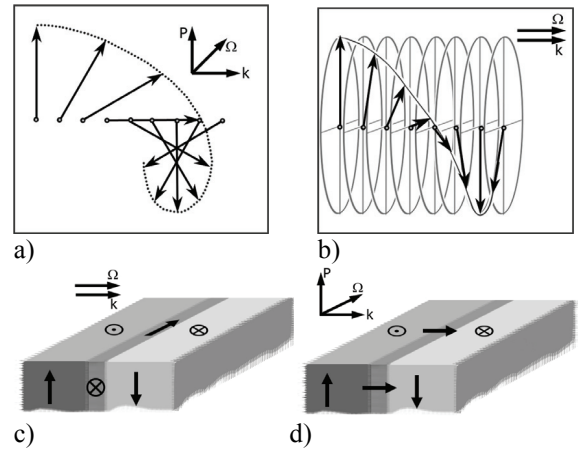

Figure 2. Schematic representation of spatially modulated spin structures and domain walls analogous to them: a) a spin cycloid b) a spin helicoid structure) the Neel-type domain wall d) the Bloch-type domain wall.

In the highly symmetrical case of a crystal with cubic symmetry (1) takes an elegant form:

$$F_{ME} = \gamma \cdot \left( \mathbf{P} \cdot \{ \mathbf{n} \cdot div(\mathbf{n}) + [\mathbf{n} \times curl(\mathbf{n})] \} \right), \quad (2)$$

where $\mathbf{n}$ is the order parameter [2]. It's noteworthy that (2) is a universal relation and describes the flexoelectric effect not only in magnets but in any ordered media, for example in liquid crystals [2]. In the last case $\mathbf{n}$ stands for director.

If we introduce the unit vector of spatial spin modulation **k** then $\nabla$ can be represented as $\mathbf{k}\partial/\partial x$, where *x* is the coordinate along the axis parallel to **k**, and we can rewrite (2) as

$$F_{ME} = -\gamma(\mathbf{P}\cdot[\mathbf{k}\times\mathbf{\Omega}]), \quad (3)$$

where **Ω** is the spin rotation axis. The electric polarization induced by the spatial spin modulation is proportional a vector product of k and Ω:

$$\mathbf{P} = -\partial F_{ME}/\partial \mathbf{E} = \gamma\chi_e[\mathbf{k}\times\mathbf{\Omega}], \quad (4)$$

where $\chi_e$ is the electric susceptibility.

---

[2] It should be noted however that in low symmetry crystal the proper-screw type of magnetic spiral can also induce electric polarization (for example, in case of crystal whose lattice structure lacks the 2-fold axis perpendicular to the direction of spin modulation [25]).

According to the simple rule (4) proposed in [10] the switching of the spiral vector chirality, i.e the direction of magnetization rotation (**Ω** $\Rightarrow$ - **Ω**) results in the reversal of electric polarization (**P**$\Rightarrow$ - **P**). The ferroelectricity induced by spatially modulated spin structures were observed in orthorhombic manganites $RMnO_3$ (R=Dy, Tb) [11-14,19], $MnWO_4$ [18, 20], hexaferrites [16; 21-23] and others spiral multiferroics. The most vivid example of the electric polarization generated by a magnetic spiral was the observation of ferroelectric domains corresponding to the spin spirals with opposite vector chirality [18].

## 3. Electric polarization of magnetic domain walls

The concept of spin flexoelectricity can be extrapolated to the area of conventional micromagnetic structures. For example the domain wall of Neel type is a soliton-like solution for a cycloid structure (fig.2c) and thus can have electrical polarization while the Bloch-type wall (fig. 2d) is an analogue for a helicoid with zero electric polarization (**Ω** || **k**).

It is worth mentioning that the term "inhomogeneous magnetoelectric interaction" was originally coined by V.G. Bar'yakhtar [1] in the context of the problem of domain wall ferroelectricity in magnets. This topic was revisited in theoretical papers [3,10,26,27]. Despite the fact that magnetoelectricity of domain walls could in principle appear in every magnetic dielectric up to the present time the experimental proof for it has remained scarce. The enhancement of electric field induced Faraday rotation of light polarization (electro-magnetooptical effect [28]) in the vicinity of domain wall observed in yttrium iron garnet films can serve only as an indirect evidence [29,30].

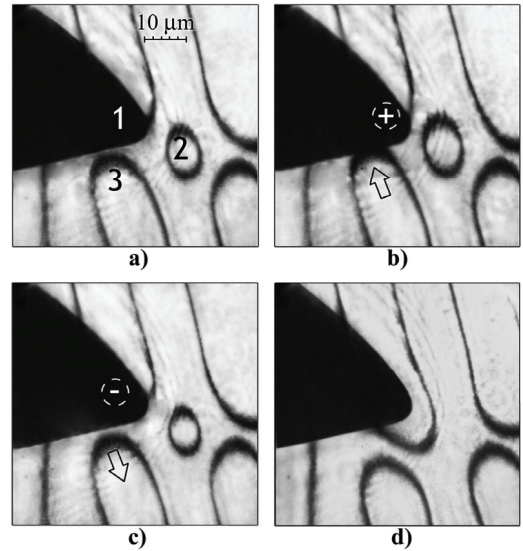

Figure 3. The electric field induced micromagnetic structure transformation. a) Initial state with no voltage applied: 1 is the image of the tip, 2 is a bubble domain, 3 is a domain head. b) displacement of the stripe domain head and bubble domain nearest to the tip-sample contact towards the tip at electric potential +500V at the tip, c) the domain walls displacement at negative potential –500 V at the tip. d) the irreversible changes of the micromagnetic structure was also observed

In our experiment we have directly observed the motion of domain walls in the gradient electric field provided by a tip electrode [31-34]. Figure 3 shows the magnetooptical images of the micromagnetic structure transformation under the influence of electric field in a $(BiLu)_3(FeGa)_5O_{12}$ iron garnet film epitaxially grown on (210) $Gd_3Ga_5O_{12}$ substrate. It can be seen as the stripe domain head displacement and inflating of bubble domain (Fig. 3 a,b). As soon as the DC voltage was switched off the domain walls came back to the equilibrium positions. Reversing the polarity of the voltage caused the opposite changes in micromagnetic structure (Fig. 3c). The repulsion (Fig. 3c) is not as evident as the attraction (fig. 3b) due to the inhomogeneity of electric field from the tip. The latter fact can be used to reduce the control voltages by scaling down the tip curvature radius. Finally, the irreversible transformation of micromagnetic structure was also observed with the coalescence of bubble domain and domain head (Fig.3 d).

The puzzling feature of the observed effect was that every domain wall has the same polarity irrespective of its position or shape, i.e. it always attracts to the positively charged tip and repels from the negatively charged one. This was proved for all the samples studied in [32]. In the context of spin flexoelectricity it can be interpreted as follows: the sense of magnetization rotation is the same for every domain wall. This fact can be explained by violation of the central symmetry in the magnetic film. The evidence for this inversion symmetry breaking in iron garnet films can be found in the paper on the linear electro-magnetooptical effect [28]. So the spontaneous state of iron garnet films is the "monochiral" micromagnetic structure and domain wall have the same electric polarization as if it were some kind of electric field built-in the film. So we were looking for the way to control the polarization of the wall and the force that could compete with this inherent mechanism.

We have found that external magnetic field perpendicular to the domain wall plane induces the phase transition from the spontaneous "monochiral" state to the state with opposite sense of magnetization rotation in the neighboring domain walls [34]. The external magnetic field reorients magnetization in the centers of the walls thus imposing alternating sense of magnetization rotation in the neighbouring domain walls (fig. 4). In accordance to (4) that means the opposite surface charges shown at the domain walls images in figure 4 are induced. The reversal of the in-plane magnetic field results in switching of the chirality and the electric polarity (fig 4 c,d).

These results agree with formula (4) as well as the theoretical model [26] and symmetry analysis [27] carried out for domain wall in electric and magnetic field applied simultaneously.

The direct coupling between electric polarity of the domain wall and its chirality evident in these experiments enables us to rule out the other possible mechanisms of domain wall magnetoelectricity not related to the spin spiral such as electric field induced anisotropy variation [29] or $P^2M^2$ magnetoelectric coupling [35]. Spin flexoelectricity should not also be confused with the similar phenomenon of electrically charged magnetic textures in magnetic film on the surface of topological insulator that was predicted in [36]. In that case the electric charges would couple not to the chirality of the spiral but to the magnetic charges ($\nabla \cdot \mathbf{m}$) [36].

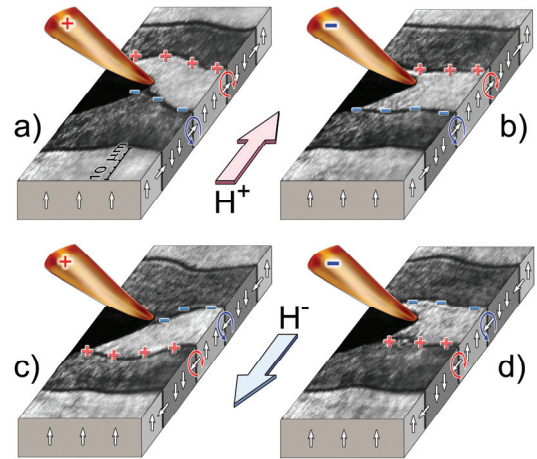

Figure 4. (Color online) The dependence of the domain walls electric polarity on its micromagnetic structure that is transformed by an external magnetic field perpendicular to the domain wall. Four cases corresponding to H =±50 Oe and two signs of electric potential of the tip V = ±1 kV are presented as the combinations of the magneto-optical image (top layer) and schematic picture of the micromagnetic configuration (cross-sections). The clockwise rotation of the magnetization in the wall corresponds to the upward electric polarization and positive surface charges shown with "+"; the counterclockwise rotation corresponds to the negative charges shown by "−" in the pictures.

## 4. Electric field control of magnetic vortexes

Micromagnetic structures are not restricted to the domain walls only. Magnetic vortex states that can be stabilized in magnetic nanodots and their dynamics are also extensively studied nowadays. This interest is largely caused by further reduction of magnetic memory cell size and introduction of patterned media. The control of vortex characteristics such as the chirality (the circulation of magnetization) [37] and the polarity of the core (the direction of the magnetization in the center of the vortex) [38] have been proposed. However all this techniques imply high current density and energy losses. There is an urgent need for a current-free vortex control.

Magnetic vortex as magnetic inhomogeneity induces bound charges in the vortex core region [10]. That makes possible the control of the third characteristic of magnetic vortex, i.e. the winding number (fig.5): vortex state can be switched to antivortex one by electric field [39].

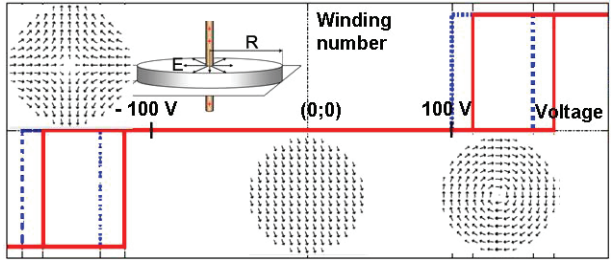

Figure 5 (color online) The hysteresis cycle of vortex/antivortex switching in electric field. The insets show (from left to right) the micromagnetic distributions of antivortex, homogeneous and vortex states. The solid line is hysteresis for particle with $M_S$ =5G, the dotted line is for 50G.

We have conducted our micromagnetic simulation using micromagnetic package SpinPM [40], which was modified to include the spin flexoelectric contribution as the effective magnetic field:

$$H_i^{eff} = -\partial F_{ME}/\partial M_i =$$
$$= (\gamma/M_s)\left[2(E_i \nabla_j m_j - E_j \nabla_i m_j) - m_j(\nabla_i E_j - \nabla_j E_i)\right] \quad (5)$$

Along with effective fields of exchange, magnetic anisotropy, and resultant field of magnetostatic dipole-dipole interaction the term (5) is the contribution to the Landau-Lifshitz-Gilbert equation for magnetization dynamics [41]. For the case of static fields only the first of the summands in (5) remains.

Figures 5 presents the results of numerical simulations of hysteresis cycle in electric field: nucleation, stabilization and disappearance of the vortex and antivortex states of 100-nm – diameter magnetic particle. The material parameters typical for magnetic dielectrics with high magnetic ordering temperature was chosen ($M_s$=5-50G, magnetoelectric constant $\gamma=10^{-6}$ (erg/cm)$^{1/2}$, exchange constant A=3·10$^{-7}$ erg/cm). The electric field was supposed to be produced by the wire with 5-nm radius running along the Z axis through the center of the particle.

The absolute value of electric field magnitude needed to nucleate antivortex is higher than that is needed for vortex creation. It is quite natural because the antivortex state costs higher magnetostatic energy. In materials with lower saturation magnetization (and lower stray fields), the picture becomes more symmetrical (fig. 5, solid line).

The problem of high control voltages still remains in the case of magnetic vortex switching. However if the magnetostatic interaction is nearly equal to the exchange interaction the two metastable states corresponding to vortex and homogeneous magnetization can be obtained in low voltage region [42].

Of special interest are vertical Bloch lines, i.e. the inhomogeneities in the domain walls that can nucleated in Bloch domain wall by magnetic field or thermal heating by focused laser light [43]. Due to its intricate micromagnetic structure the volume and surface density of charges was expected to develop in it (fig.6, for details, see [44]). The electric properties of vertical Bloch lines have been observed as electric field induced displacement of Bloch lines along the domain walls [44].

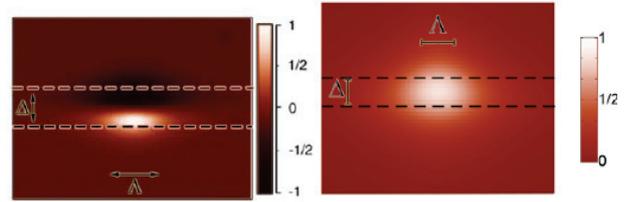

Figure 6. (Color online) the top view of a) volume b)surface electric charge distribution in the domain wall with the vertical Bloch line (results of numerical simulations). The domain wall is shown with dashed line, Δ is the domain wall width, Λ is vertical Bloch line width.

## 5. Conclusion

The inhomogeneous magnetization distribution on the micro and nanoscale locally reduces the symmetry of the magnetic crystal and can induce the accompanying electric polarization distribution. In the presence of external electric field (or the internal one associated with spontaneous polarization) the spin flexoelectricity can be an important factor of micromagnetism like exchange, magnetostatic interaction or magnetic anisotropy.

Besides its fundamental importance the electric charge density associated with magnetic inhomogeneities provides new means for the electrical control of micromagnetic structure, e.g. domain wall motion triggered by electric field and electrical control of magnetic vortices in magnets. This fits the trends in low-power-consumption spintronics and magnetic memory. Low spin damping in iron garnet films compared to multiferroic materials makes them also interesting in the context of electrically tuned spin wave propagation.

**Acknowledgements** Authors are grateful to A.V. Nikolaev, E.P. Nikolaeva, D.A. Sechin, A.S. Sergeev for cooperation. In this paper we also commemorate Prof. A.S. Logginov (1940-2011) impact on experimental research in this field. The support from RFBR grant № 11-02-12170-ofi-m-2011 and 10-02-13302-RT-omi is acknowledged.

**References**

[1] V. G. Bar'yakhtar, V. A. L'vov, and D. A. Yablonskii, JETP Lett. 37 (1983) 673.
[2] A. Sparavigna, A. Strigazzi, and A. Zvezdin, Phys. Rev. B 50 (1994) 2953.
[3] I. Dzyaloshinskii, EPL 83 (2008) 67001.
[4] D. L. Mills and I. E. Dzyaloshinskii, Phys. Rev. B 78 (2008) 184422.
[5] A.A. Mukhin, A. K. Zvezdin, JETP Lett., 89 (2009) 328.
[6] A.P. Pyatakov, A.K. Zvezdin, The European Physical Journal B - Condensed Matter and Complex Systems 71 (2009) 419.
[7] R. E. Newnham, J. J. Kramer, W. A. Schulze, L. E. Cross, J. Appl. Phys., 49 (1978) 6088.
[8] G. A. Smolenskii and I. Chupis, Sov. Phys. Usp. 25 (1982) 475.
[9] Yu.F. Popov, A.K. Zvezdin, G.P. Vorob'ev, A.M. Kadomtseva, V.A. Murashev, D. N. Rakov, JETP Lett., 57 (1993) 69.
[10] M. Mostovoy, Phys. Rev. Lett. 96 (2006) 067601.
[11] T. Kimura, T. Goto, H. Shintani, K. Ishizaka, T. Arima, Y. Tokura, Nature 426 (2003) 55.
[12] T. Kimura, Annu. Rev. Mater. Res. 37 (2007) 387.
[13] E. Milov, A. Kadomtseva, G. Vorob'ev, Y. Popov, V. Ivanov, A. Mukhin, A. Balbashov, JETP Lett. 85 (2007) 503.
[14] Y. Yamasaki, H. Sagayama, T. Goto, M. Matsuura, K. Hirota, T. Arima, Y. Tokura, Phys. Rev. Lett. 98 (2007) 147204.
[15] S.-W. Cheong, M. Mostovoy, Nature Materials 6 (2007) 13.
[16] Sh. Ishiwata, Y. Taguchi, H. Murakawa, Y. Onose, and Y. Tokura, Science 319 (2008)1643.
[17] V. V. Men'shenin, JETP Volume 108 (2009) 236.
[18] D. Meier, M. Maringer, Th. Lottermoser, P. Becker, L. Bohatý, and M. Fiebig, Phys. Rev. Lett. 102 (2009) 107202
[19] F. Kagawa, M. Mochizuki, Y. Onose, H. Murakawa, Y. Kaneko, N. Furukawa, and Y. Tokura, PRL 102 (2009) 057604.
[20] T. Finger, D. Senff, K. Schmalzl, W. Schmidt, L. P. Regnault, P. Becker, L. Bohatỳ, and M. Braden, Phys. Rev. B 81 (2010) 054430.
[21] S. H. Chun et al, PRL 104 (2010) 037204.
[22] Y. Kitagawa, Y. Hiraoka, T. Honda, T. Ishikura, H. Nakamura, Ts.Kimura, Nature Materials, 9, 797;
[23] M. Soda et al PRL 106 (2011) 087201.
[24] Y. Tokura, Sh. Seki, Advanced Materials 22 (2011)1554.
[25] Taka-hisa Arima, J. Phys. Soc. Jpn. 76 (2007) 073702.
[26] A.A. Khalfina, M.A. Shamtsutdinov, Ferroelectrics 279 (2002) 19.
[27] Tanygin B., J. Magn. & Magn. Mater. 323 (2011) 616.
[28] B. B. Krichevtsov, V. V. Pavlov, and R. V. Pisarev, JETP Lett. 49 (1989) 535.
[29] V.E. Koronovskyy, S.M. Ryabchenko, V.F. Kovalenko, Phys. Rev. B 71 (2005) 172402.
[30] V.E. Koronovskyy, Appl Phys A 95 (2009) 351.
[31] A.S. Logginov, G.A. Meshkov, A.V. Nikolaev, and A.P. Pyatakov, JETP Letters 86 (2007) 115.
[32] A.S. Logginov, G.A. Meshkov, A.V. Nikolaev, E.P. Nikolaeva, A.P. Pyatakov, A.K. Zvezdin Applied Physics Letters 93 (2008) 182510.
[33] Logginov A., Meshkov G., Nikolaev A., Nikolaeva E., Pyatakov A., Zvezdin A., Solid State Phenomena 152-153 (2009) 143
[34] A. P. Pyatakov, D. A. Sechin, A. S. Sergeev, A. V. Nikolaev, E.P. Nikolaeva, A. S. Logginov and A. K. Zvezdin, EPL 93 (2011) 17001.
[35] Daraktchiev M., Catalan G. and Scott J. F., Phys. Rev. B, 81 (2010) 224118.
[36] Nomura K. and Nagaosa N., Phys. Rev. B, 82 (2010) 161401(R).
[37] M. Tanase, A. K. Petford-Long, O. Heinonen, K. S. Buchanan, J. Sort, and J. Nogu´es, Phys. Rev. B, 79 (2009) 014436.
[38] K. Yamada, Sh. Kasai, Y. Nakatani, K. Kobayashi, H. Kohno, A.Thiaville, T. Ono. Nat Mater, 6 (2007) 270–273.
[39] A. P. Pyatakov, G. A. Meshkov, and A. S. Logginov, Moscow University Physics Bulletin, 65(4) (2010) 329–331.
[40] A. V. Khvalkovskiy, J. Grollier, N. Locatelli, Ya. V. Gorbunov, K. A. Zvezdin, and V. Cros, Applied Physics Letters 96 (2010) 212507.
[41] Th. Fischbacher, M. Franchin, and H. Fangohr, Journal of Applied Physics 109(7) (2011) 07D352.
[42] G.A. Meshkov, A.P. Pyatakov, A.D. Belanovsky, K.A. Zvezdin, A.S. Logginov, J. Magn. Soc. Japan, 36 (1–2) (2012) 46–48.
[43] A.S. Logginov, A.V. Nikolaev, JETP Lett. 66 (1997) 426.
[44] A.S. Logginov, G.A. Meshkov, A.V. Nikolaev, A. P. Pyatakov, V. A. Shust, A.G. Zhdanov, A.K. Zvezdin, Journal of Magnetism and Magnetic Materials, 310 (2007) 2569.